\documentclass[10pt, twocolumn, conference]{IEEEtran}

\usepackage{graphicx,epsfig,color,bm}
\usepackage{color}
\usepackage{amsmath,epsfig} % easier math formulae, align, subequation
\usepackage{times}
\usepackage{enumerate}
\usepackage{amsfonts}
\usepackage{bm}
\usepackage{amssymb}
\usepackage{cite}

\newcommand{\beq}{\begin{equation}}
\newcommand{\eeq}{\end{equation}}
\newcommand{\ben}{\begin{enumerate}}
\newcommand{\een}{\end{enumerate}}
\newcommand{\ber}{\begin{eqnarray}}
\newcommand{\eer}{\end{eqnarray}}

\newcommand{\beqa}{\begin{eqnarray}}
\newcommand{\eeqa}{\end{eqnarray}}

\newtheorem{thm}{Theorem}

\newtheorem{prop}{Proposition}

\newtheorem{defn}{Definition}

\newcommand{\abs}[1]{\left\vert#1\right\vert}
\newcommand{\norm}[1]{\Vert#1\Vert}

\title{Caching with Unknown Popularity Profiles in Small Cell Networks}
\author{
\authorblockN{Bharath Bettagere Nagaraja}
\authorblockA{Department of ECE,\\ PESIT-Bangalore South Campus,\\
Bangalore, India.\\
Email: \texttt{bharathbn@pes.edu}}
\and
\authorblockN{Kyatsandra G. Nagananda}
\authorblockA{Department of ECE,\\ PES University,\\
Bangalore, India.\\
Email: \texttt{kgnagananda@pes.edu}}
}
\begin{document}

\maketitle
\thispagestyle{empty}
\pagestyle{empty}

\begin{abstract}
A heterogenous network is considered where the base stations (BSs), small base stations (SBSs) and users are distributed according to independent Poisson point processes (PPPs). We let the SBS nodes to posses high storage capacity and are assumed to form a distributed caching network. Popular data files are stored in the local cache of SBS, so that users can download the desired files from one of the SBS in the vicinity subject to availability. The offloading-loss is captured via a cost function that depends on a random caching strategy proposed in this paper. The cost function depends on the popularity profile, which is, in general, unknown. In this work, the popularity profile is estimated at the BS using the available instantaneous demands from the users in a time interval $[0,\tau]$. This is then used to find an estimate of the cost function from which the optimal random caching strategy is devised. The main results of this work are the following: First it is shown that the waiting time $\tau$ to achieve an $\epsilon>0$ difference between the achieved and optimal costs is finite, provided the user density is greater than a predefined threshold. In this case, $\tau$ is shown to scale as $N^2$, where $N$ is the support of the popularity profile. Secondly, a transfer learning-based approach is proposed to obtain an estimate of the popularity profile used to compute the empirical cost function. A condition is derived under which the proposed transfer learning-based approach performs better than the random caching strategy.
\end{abstract}

\section{Introduction} \label{sec:intorduction} \vspace{-0.05in}
The advent of multimedia-capable devices such as smart phones at economical costs has triggered the growth of mobile data traffic at an unprecedented rate. For instance, computationally intensive applications like content-based video analysis alone contributes a large fraction of the total mobile data traffic. This trend is likely to continue at an alarming pace, demanding wireless service providers to reevaluate the design strategies for the next generation wireless infrastructure - providing high data rates for tens of thousands of users, improving the spectral efficiency beyond what is currently available, improving coverage and cell-capacity, and engineering efficient inter-device signalling techniques \cite{Akyildiz2004}.

\vspace{0.015in}
A promising approach to address the wireless infrastructure problem is to deploy small cells which can offload a significant amount of data from the macro BS \cite{Chou2014}. Research on next generation wireless infrastructure has primarily focussed on the design and implementation of such small cells \cite{Bennis2013}. However, a shortcoming of small cell-based architectures is that, during peak traffic hours, the backhaul-link capacity requirement to support data traffic is enormously high \cite{Kim2014}. At the same time, the cost incurred in designing high capacity backbone network for small cells is quite discouraging especially from an economic standpoint. Thus, in the grand scheme of things, small cell-based solutions alone will not suffice to efficiently address the quality of service requirements associated with future traffic demands. Therefore, one is inspired to further investigate the advantages offered by heterogenous networks.

A recent development in this direction is to improve the accessibility of data content to users by storing the most popular data files in the \emph{local caches} of small cell BSs, with the objective of reducing the peak traffic rates \cite{Lin2003} - \nocite{Bastug2014} \nocite{Niesen2012}\cite{Sadjadpour2014}. This is commonly referred to as ``caching'' and has attracted significant attention. For instance, in \cite{Ali_Niessen_ISIT_may2014}, it was shown that, when  cached-content demand was uniformly distributed, joint optimization of caching and coded multicast delivery significantly improved the caching gains; this was extended to a more general decentralized setting in \cite{Maddah-Ali2014a}. The problem setup considered in \cite{Ali_Niessen_ISIT_may2014} was extended to nonuniform distribution on cached-content demand in \cite{Niessen_Ali_INFOCOM_April2014}. The tradeoff between the performance gain of coded caching and delivery delay in computationally intensive applications such as video streaming was characterized in \cite{Niessen_Ali_arxiv_2014}, \cite{Pedarsani_Ali_INFOCOM_ICC_June2014}; while, in \cite{Karamchandani_Niesen_Ali_Diggavi_arxiv_2014}, coded caching was achieved for content delivery networks with two layers of caches. A polynomial-time heuristic solution was proposed in \cite{Guan2014} to address the NP-hard optimization problem of maximizing the caching utility of users based on the mobility patterns of users.

Caching has also made advances in device-to-device (D2D) communications. In \cite{Ji_Caire_Molisch_ISIT_July2013}, the outage-throughput tradeoff was characterized for D2D nodes having access to cached files in a library, but obtained the requested file from a third-party via a D2D link. In \cite{Golrezaei_Dimakis_Molisch_ISIT_July2013}, the conflict between collaboration-distance and interference was identified among D2D nodes to maximize frequency reuse by exploiting distributed storage of cached content in those nodes. In \cite{Ji_Caire_Molisch_ITW_Sep2013}, coded caching was shown to achieve multicast gain in a D2D network, where users had access to linear combinations of packets from cached files. The NP-hard nature of distributed caching was reported in \cite{Golrezaei_Shanmugam_Dimakis_Molisch_TIT_Dec2013}, where approximation algorithms were proposed for video content delivery. In \cite{MingyueJi_Caire_Molisch_arxiv_2014}, the throughput scaling laws of random caching, where users with pre-cached information made arbitrary requests for cached files, was studied.
\vspace{-0.05in}
\subsection{Main contributions of this paper}\label{subsec:main_contributions}%\vspace{-0.05in}
In the above cited references, the popularity profile of data files was assumed to be known perfectly. However, in practical applications, this turns out to be a very weak assumption. In this paper, we relax the assumption on the knowledge of the popularity profile to devise a caching strategy. %An overview of the problem setup considered in this paper and the solution methodology is provided in the following paragraph.

We consider a heterogenous network where the users, BS and SBSs are assumed to be distributed according to independent PPPs. Each SBS is assumed to employ a random caching strategy with no caching at the user terminal (see \cite{Ji_Caire_Molisch_ISIT_July2013}). A protocol model for communications is proposed using which a cost, which captures the backhaul-link overhead that depends on the popularity profile, is derived (Section \ref{sec:sys_model}). Secondly, assuming a Poisson request model, a centralized approach is presented in which BS computes \textbf{an estimate of the popularity profile}; this estimate is then used in the cost function to optimize the caching probability. Thus, the actual cost incurred differs from the optimal cost, and the difference depends on the number of samples used to estimate the popularity profile. Furthermore, the number of samples collected at BS is related to the density of the Poisson arrival process and the waiting time during which the samples are collected. We derive a lower bound on this waiting time which guarantees a cost that is within $\epsilon > 0$ of the optimal cost (this is derived in Theorem \ref{thm:time_complexity_centralized}, Section \ref{sec:target_domain_only_learning}).

The following are the main findings of our study:
\begin{enumerate}[1)]
\item The waiting time is finite, provided the user density is greater than a threshold.
\item The waiting time scales as $N^2$, where $N$ is the total number of cached data files in the system.
\item These results are improved by using a transfer learning (TL)-based approach to estimate the popularity profile wherein samples from other domains such as those obtained from a social network are used to improve the estimation accuracy. The minimum number of source domain samples required to achieve a better performance is derived. Employing the TL-based approach, a finite waiting time is achieved for all user density.
\item In the TL-based approach, the waiting time is a function of the ``distance'' between the probability distribution of the files requested and the distribution of the source domain samples (the notion of distance is made precise in the proof of Theorem \ref{thm:time_complexity_centralized_TL}, Section \ref{sec:transfer_learning}).
\end{enumerate}
Our work draws parallels with those reported in \cite{Blasco_Gunduz_ICC_June2014} - \nocite{Blasco_Gunduz_ISIT_June2014}\cite{Sengupta_Amura_Tandom_Buehrer_Clancy_ISWCS_2014}, where a learning-based approach was used in which the popularity profile was estimated to devise the caching strategy. We emphasize that, unlike \cite{Blasco_Gunduz_ICC_June2014} - \nocite{Blasco_Gunduz_ISIT_June2014}\cite{Sengupta_Amura_Tandom_Buehrer_Clancy_ISWCS_2014}, this paper focuses on theoretical analysis of time and sample complexities to  achieve the desired performance. To the best of our knowledge, this is first instance where an analytical treatment of waiting time and its relation to the probability distribution function of source domain samples has appeared in the literature on caching.

The rest of the paper is organized as follows. In Section \ref{sec:sys_model}, the system model is first presented followed by a communications protocol for caching. The two methods for estimating the popularity profile and its corresponding waiting time analysis are developed in Section \ref{sec:learning_complexity}. Concluding remarks are provided in Section \ref{sec:conclude}. The proofs of main theorems of the paper are relegated to appendices.

%The following notation is used throughout the paper. The shorthand $\{x\}^+ \triangleq \max\{x,0\}$. Random variables are denoted by upper case letters, while the realizations of random variables are denoted by lower case letters. The expected value of $X$ is denoted by $\mathbb{E}\{X\}$ and the expectation of $X$ conditioned on $Y$ is denoted $\mathbb{E} \left\{X | Y\right\}$. The indicator function $\mathbf{1}\{A\}$ is equal to one if event $A$ occurs, and zero otherwise.

\section{System Model and Communications Protocol} \label{sec:sys_model}
A heterogenous cellular network is considered where the set of users (denoted $\Phi_u \subseteq \mathbb{R}^2$), the set of BSs (denoted $\Phi_b \subseteq \mathbb{R}^2$), and the set of SBSs (denoted $\Phi_s \subseteq \mathbb{R}^2$) are distributed according to independent PPP with density $\lambda_s$, $\lambda_b$ and $\lambda_u$, respectively \cite{FBaccelli_SpringerJS_1997}. Each user independently requests for a data file of size $B$ bits in $\mathcal{F}\triangleq \{f_1,f_2,\ldots,f_N\}$ with popularity $\mathcal{P}\triangleq \{p_1,\ldots,p_N\}$, $\sum_{i=1}^N p_i = 1$, and is assumed to be stationery across time. In a typical heterogenous cellular network, BS fetches the file using its backhaul link to serve the user. During peak data traffic hours, this results in an information-bottleneck both at BS as well as in its backhaul link. To overcome this problem, caching of the most popular files (either at the user nodes or at SBSs) is proposed -- the requested file will be served directly by one of the neighboring SBS depending on the availability of the file in its local cache. The performance of caching depends on the density of SBS nodes, the cache size, the users' request rate, and the caching strategy. It is assumed that the SBS can cache up to $M$ files, each of length $B$ bits. Each SBS $s \in \Phi_s$ caches its content in an i.i.d. manner by generating $M$ indices distributed according to $\Pi \triangleq \{\pi_i: f_i \in \mathcal{F}\}$, $\sum_{i=1}^N \pi_i = 1$ (see \cite{Ji_Caire_Molisch_ISIT_July2013}).

The following communication protocol is used in this paper which determines the set of neighbor SBS nodes for any user $u$. Each SBS $s$ at $x_s \in \Phi_s$ communicates with a user $u$ at $x_u  \in \Phi_u$ if $\norm{x_u - x_s} < \gamma$, ($\gamma > 0$); this condition determines the communication radius. Note that, we have ignored the interference constraint. The set of potential neighbors of user $u$ at $x_u$ is denoted $\mathcal{N}_u \triangleq \{y \in \Phi_s: \norm{y - x_u} < \gamma\}$.

The performance of the caching scheme depends on the metric used to optimize the strategy. We focus on the performance of a typical user located at the origin. Since the main goal is to reduce the information-bottleneck at BS, the objective is to minimize the time overhead due to the unavailability of the requested file.
%Moreover, even if the requested file is present in the neighborhood of the user, the interference constraints may not be satisfied for $T>0$ consecutive time slots. This can exceed the delay tolerance of the node, and the node may request BS to transmit the entire file. Following these observations,  the transmission model is defined as follows.
%\defn (\emph{Transmission Model}) Each user $u \in \Phi_u$ independently transmits with probability $q$.
%\defn ($T$-\empty{feasible})
%A node $u \in \Phi_u$ is said to be $T$-feasible if there exists at least one in $T$ consecutive time slots in which there does not exists a transmitting node in a $\zeta$ neighborhood of the user.
The following ``offloading loss" metric is used to optimize the caching strategy proposed in this paper:
\beq
\mathcal{T}(\Pi,\mathcal{P})  \triangleq \frac{B}{R_0}\mathbb{E} \left[\mathbf{1}\{f_i \notin \mathcal{N}_u\} \right],\hspace{-2mm}
%&=&\hspace{-2mm} \frac{B}{R_0}\mathbb{E} \left[\mathbf{1}\{f_i \notin \mathcal{N}_u\} \}\right],
\label{eq:metric}
\eeq
%where $\mathcal{E}:= \{u \text{ is $T$-feasible}\}$.
which captures the unavailability of the requested file from a typical user at the origin. $R_0$ is the rate supported by BS to the user at the origin, and $\frac{B}{R_0}$ is the time overhead incurred in transmitting the file from BS to the user. The expectation is with respect to $\Phi_u$, $\Phi_s$ and $\mathcal{P}$. The indicator function $\mathbf{1}\{A\}$ is equal to one if event $A$ occurs, and zero otherwise. Better performance can be achieved by minimizing $\mathcal{T}(\Pi,\mathcal{P})$, i.e.,
\beqa \label{eq:opt_problem}
\min_{\Pi \succeq 0} ~~ \mathcal{T}(\Pi,\mathcal{P}) \\ \nonumber
\text{       subject to } \sum_{i=1}^N \pi_i = 1, ~\pi_i \geq 0,
\eeqa
for $i=1,\ldots,N$. To solve the optimization setup \eqref{eq:opt_problem}, we need an expression for $\mathcal{T}(\Pi,\mathcal{P})$, which is derived in the following.

\begin{thm}  \label{thm_mean_throughput}
For the caching strategy proposed in this paper, the average offloading loss is given by
\beqa \label{eq:mean_througput_expression}
\mathcal{T}(\Pi,\mathcal{P})  = \frac{B}{R_0} \left[ \sum_{i=1}^N \exp\{-\lambda_u \pi \gamma^2(1-\pi_i)^M\} p_i \right].
\eeqa
%where $\mathcal{G}(\lambda_u, q, \gamma):= 1 - \exp\{-\lambda_u \pi \gamma^2 (1-q^T)\}$.
\end{thm}

\emph{Proof:} See Appendix \ref{app:througput_derivation}. $\blacksquare$

In general, the optimization problem \eqref{eq:opt_problem} is non-convex. More importantly, the popularity profile $\mathcal{P}$ is unknown, and is estimated from the available data. Denoting the estimated popularity profile by $\hat{\mathcal{P}} \triangleq \{\hat{p}_1,\ldots,\hat{p}_N\}$ and the corresponding offloading loss by ${\mathcal{T}}(\Pi,\hat{\mathcal{P}})$, problem \eqref{eq:opt_problem} can be reformulated as follows:
\beqa
\min_{\Pi \succeq 0} ~~ {\mathcal{T}}(\Pi,\hat{\mathcal{P}}) \label{eq:opt_problem_emperical} \\ \nonumber
\text{subject to} \sum_{i=1}^N \pi_i = 1.
\eeqa

The solution to the reformulated problem \eqref{eq:opt_problem_emperical} differs from that of the original problem \eqref{eq:opt_problem}. Let $\hat{\Pi}^*$ and $\Pi^*$ denote the optimal solutions to the problems in \eqref{eq:opt_problem} and \eqref{eq:opt_problem_emperical}, respectively, and let the throughput achieved using $\hat{\Pi}^*$ be denoted $\hat{\mathcal{T}}^* \triangleq \mathcal{T}(\hat{\Pi}^*,{\mathcal{P}})$. The central theme of this paper is the analysis of the offloading loss difference $\hat{\mathcal{T}}^* - \mathcal{T}^*$, where $\mathcal{T}^* \triangleq \mathcal{T}(\Pi^*, \mathcal{P})$. Theorem \ref{thm:time_complexity_centralized} and Theorem \ref{thm:time_complexity_centralized_TL} are given over to this analysis. We assume the minimization problems \eqref{eq:opt_problem} and \eqref{eq:opt_problem_emperical} can be solved in polynomial time.

\section{Learning Complexity} \label{sec:learning_complexity}
In this section, we study two quantities of interest, namely, the time complexity involved in obtaining the samples and estimating the empirical risk to achieve a performance within $\epsilon>0$ of the optimal solution. The efficiency of the estimate $\hat{\mathcal{P}}$ of the popularity profile depends on the number of available samples, which in turn is related to the number of requests made by the users. We first define the user-request model.

\defn (\emph{Request Model}) Each user requests for a file $f \in \mathcal{F}$ at a random time $t \in [0,\infty]$ following an independent Poisson arrival process with density $\lambda_r>0$.

For notational convenience, the same density is assumed across all the users. A centralized scheme is used where BS collects the requests from all the users in its coverage area in the time interval $[0,\tau]$, to estimate the popularity profile of the requested files. Let the number of users in the coverage area of BS $b \in \Phi_b$ of radius $R>0$ be $n_R$, which is distributed according to a PPP with density $\lambda_u$. Let the number of requests made by the user $u \in \{\Phi_u \bigcap \mathbb{B}(0,R)\}$ in the time interval $[0,\tau]$ be $k_u$, where $\mathbb{B}(0,R)$ is a two-dimensional ball of radius $R$ centered at $0$. We assume that requests across the users are known at BS. The requests from the user $u$ is denoted $\mathcal{X}_u \triangleq \{X_u^{(1)},\ldots,X_u^{(k_u)}\}$, where $X_u^{(l)} \in \{1,\ldots,N\}$ denotes the indices of the files in $\mathcal{F}$, $l=0,\ldots,k_u$. After receiving $\mathcal{X}_u$, $u\in \{\Phi_u \bigcap \mathbb{B}(0,R)\}$, in the time interval $[0,\tau]$, BS computes an estimate of the popularity profile as follows:
\beq \label{eq:estimation_popularity}
\hat{p}_i = \frac{1}{\sum_{u \in \{\mathbb{B}(0,R) \bigcap \Phi_u\}}  k_u} \sum_{l=0}^{k_u}{\sum_{u \in \mathbb{B}(0,R) \bigcap \Phi_u } \mathbf{1}\{X_u^{(l)} = i\}},
\eeq
$i=1,\ldots,N$. Given the number $n_R$ of users in the coverage area of BS, the term $\sum_{u \in \{\mathbb{B}(0,R) \bigcap \Phi_u\} } k_u$ is a PPP with density $n_R \lambda_r$. Also, $\mathbb{E} \left\{\hat{p}_i  |  \abs{\{\Phi_u \bigcap \mathbb{B}(0,R)\}} = n_R\right\} = p_i$, from which we conclude that $\hat{p}_i$ is an unbiased estimator. The estimated popularity profile $\hat{p}_i$ is shared with every SBS in the coverage area of BS, and then used in \eqref{eq:opt_problem_emperical} to find the optimal caching probability. The derived estimator can be improved by using samples from other related domains, for example, a social network. The term ``target domain'' is used when samples are obtained only from users in the coverage of BS. Next, we derive the minimum waiting time $\tau$ required to achieve the desired accuracy of $\epsilon$.

\subsection{Target domain-only learning} \label{sec:target_domain_only_learning}
In this subsection, we derive a lower bound on the waiting $\tau$ corresponding to the estimator in \eqref{eq:estimation_popularity}.

\begin{thm} \label{thm:time_complexity_centralized}
For any $\epsilon>0$, with a probability of at least $1-\delta$, an offloading loss of $\hat{\mathcal{T}}^* \leq \mathcal{T}^* + \epsilon$ can be achieved using the estimate in \eqref{eq:estimation_popularity} provided
\beq \label{eq:sourceonly_waitingtime}
\tau \geq \left\{ \begin{array}{cc} \left\{\frac{1}{\lambda_r g^*} \log \left(\frac{1}{1- \frac{1}{\lambda_u \pi R^2} \log \frac{2 N}{\delta}}\right)\right\}^+  &\text{if } \lambda_u > \mathcal{L},\\
\infty  &\text{otherwise},
\end{array}
\right.
\eeq
where $\{x\}^+ \triangleq \max\{x,0\}$, $g^* \triangleq (1- \exp\{-2 \bar{\epsilon}^2\})$, $\mathcal{L}\triangleq \frac{1}{\pi R^2} \log \left(\frac{2 N}{\delta}\right)$ and
\beq
\bar{\epsilon} \triangleq \frac{R_0 \epsilon}{2 B \sup_{\Pi} \sum_{i=1}^N g(\pi_i)},
\eeq
with $g(\pi_i)\triangleq \exp\{-\lambda_u \pi \gamma^2 (1-\pi_i)^M\}$.
\end{thm}
\emph{Proof}: See Appendix \ref{app:time_complexity_centralized}. $\blacksquare$

To achieve a finite waiting time that results in an accuracy of $\epsilon>0$, the user density $\lambda_u$ has to be greater than a threshold.
%It is interesting to see that even if $\lambda_r \rightarrow \infty$, the delay does not go to zero unless the user density is sufficiently high.
Further insights into \eqref{eq:sourceonly_waitingtime} are obtained by making the following approximation: $1-x\leq e^{-x}$ for all $x\geq0$. This combined with $\sup_{\Pi: \Pi \succeq 0, \mathbf{1}^T\Pi = 1} \sum_{i=1}^N g(\pi_i) \leq N$
%$\exp\{-\lambda_u \pi \gamma^2 (1-1/N)^M\}$
yields the following lower bound on $\tau$:
\beq \label{eq:source_domain_only_insight_waiting_time}
\tau \geq  \frac{2 B^2}{\pi R^2 \lambda_u \lambda_r R_0^2 \epsilon^2} N^2 \log\left(\frac{2N}{\delta}\right).
\eeq
A few observations are in order:
\begin{enumerate}[(1)]
\item The waiting time $\tau$  to achieve an $\epsilon$-offloading loss difference scales as $N^2$;
\item $\tau$ is inversely proportional to ($\lambda_u$, $\lambda_r$);
\item as the radius of coverage increases, the delay decreases as $1/R^2$; and
\item as the file size $B$ increases, the waiting time scales as $B^2$.
\end{enumerate}
The above result is a lower bound on the waiting time per request per user, since the offloading loss is derived for a given request per user. There are on an average $\lambda_r$ requests per unit time per user. Thus, to obtain the waiting time per user, the offloading loss has to be multiplied by $\lambda_r$. This amounts to replacing $\epsilon$ by $\epsilon/\lambda_r$, resulting in the following lower bound:
\beq \label{eq:source_domain_only_insight_waiting_time_per_user}
\tau \geq  \frac{2 B^2 \lambda_r}{\pi R^2 \lambda_u R_0^2 \epsilon^2} N^2 \log\left(\frac{2N}{\delta}\right).
\eeq
It is clear that the waiting time scales linearly with $\lambda_r$. Although the waiting time per user per request tends to zero as $\lambda_r \rightarrow \infty$, the waiting time per user tends to $\infty$. This is because the number of requests per unit time approaches $\infty$, and thus, a small fraction of errors results in an infinite difference in offloading loss leading to an infinite waiting time.
In the following subsection, a TL-based approach is devised to improve the waiting time.

%To get more realistic figures for the minimum waiting time, the following typical values are used (see ..).
%Let $\lambda_u = 0.2$, $\lambda_r = 0.1$, $R=1.3$ and $\epsilon=0.2$.

\subsection{Transfer learning to improve the waiting time} \label{sec:transfer_learning}
In general, the minimum waiting time required to achieve an accuracy of $\epsilon > 0$ can be very large. A recent approach to alleviate this shortcoming is to utilize the knowledge gained from users' interactions with a social community (termed ``source domain''). Specifically, by cleverly combining samples from the source domain and users' request pattern (i.e., the target domain), one can potentially reduce the waiting time. However, the accuracy is indicative of the dependency between the source and target domains. Thus, TL-based approach and its impact on the time complexity to achieve a given accuracy is of paramount importance. TL-based approach was also employed in \cite{Bastug_Bennis_Debbah_KuVS_2014} to negotiate over-fitting problems in estimating the content popularity profile matrix. However, in this paper, we are interested in finding the minimum waiting time to achieve a desired performance accuracy. Furthermore, the model considered in this paper is different from the one in \cite{Bastug_Bennis_Debbah_KuVS_2014}. The TL-based approach considered here comprises two sources from which the samples are drawn, namely, source domain and target domain. 

An estimate of the popularity profile is obtained as follows. First, using target domain samples, the following parameter is estimated at BS
\beq \label{eq:sum_estimate_target}
\hat{S}_i^{(tar)} \triangleq   \sum_{l=0}^{k_u}{\sum_{u \in \mathbb{B}(0,R) \bigcap \Phi_u } \mathbf{1}\{X_u^{(l)} = i\}},
\eeq
$i=1,\ldots,N$, similar to \eqref{eq:estimation_popularity}. We denote the source domain samples $\mathcal{X}^s \triangleq \{X_1^{s},\ldots,X_m^{s}\}$, which is drawn i.i.d from a distribution $\mathcal{Q}$. Here, $X_l^s = i (i=1,\ldots,N$) denotes that the user corresponding to the $l^{\text{th}}$ sample has requested the file $f_i$. Using this, BS computes
\beq \label{eq:source_domain_sum}
\hat{S}_{i}^{s} \triangleq \sum_{k=1}^m \mathbf{1}\{X_i^{s} = i\},~i=1,2,\ldots,N.
\eeq
Now, BS uses \eqref{eq:sum_estimate_target} and \eqref{eq:source_domain_sum} to compute an estimate of $\hat{p}_i^{(tl)}$ (the superscript $(tl)$ indicates transfer learning) given by
\beq \label{eq:estimation_TL}
\hat{p}_i^{(tl)} \triangleq \frac{\hat{S}_i^{(tar)} +  \hat{S}_{i}^{s}}{{\sum_{u \in \{\mathbb{B}(0,R) \bigcap \Phi_u\}}  k_u} + m}.
\eeq

Using the estimate \eqref{eq:estimation_TL}, we provide a bound on the time complexity in the following theorem.
\begin{thm} \label{thm:time_complexity_centralized_TL}
For any accuracy
\beq \label{eq:epsilon_error_bound}
\epsilon > \frac{2 B  \sup_\Pi \left\{\sum_{i=1}^N g(\pi_i)\right\} } {R_0} \norm{\mathcal{P} - \mathcal{Q}}_{\mathcal{H}},
\eeq
with a probability of at least $1-\delta$, an offloading loss $\hat{\mathcal{T}}^* \leq \mathcal{T}^* + \epsilon$ can be achieved using the estimate in \eqref{eq:estimation_TL} provided
\beq
\tau \geq \left\{ \begin{array}{cc} \left\{\frac{1}{\lambda_r (1-e^{-2 \epsilon_{pq}^2})} \log \left(\frac{1}{1 - \Lambda} \right)\right\}^+  &\text{if } \lambda_u > \rho,\\
\infty  &\text{otherwise},
\end{array}
\right.
\eeq
where $\rho \triangleq \frac{1}{\pi R^2} \left(\log \frac{2N}{\delta} - 2 \epsilon_{pq}^2 m\right)$, $\epsilon_{pq}\triangleq \bar \epsilon - \norm{\mathcal{P} - \mathcal{Q}}_{\mathcal{H}}$,
\beq
\Lambda \triangleq \frac{1}{\lambda_u \pi R^2} \left(\log \frac{2N}{\delta} - 2 \epsilon_{pq}^2 m\right),
%\frac{R_0 \epsilon}{2 B \mathcal{G}(\lambda_u,q,\gamma) \sup_{\Pi} \sum_{i=1}^N g(\pi_i)}
\eeq
$\bar \epsilon \triangleq \frac{R_0 {\epsilon}}{2 B \sup_\Pi \left\{\sum_{i=1}^N g(\pi_i)\right\} }$, $g(\pi_i) \triangleq \exp\{-\lambda_u \pi \gamma^2(1-\pi_i)^M\}$, and $\norm{\mathcal{P} - \mathcal{Q}}_{\mathcal{H}} \triangleq \sup_{i\in[1,N]} \abs{q_i - p_i}$.
\end{thm}
\emph{Proof:} See Appendix \ref{app:TL_time_complexity}. $\blacksquare$

From Theorem \ref{thm:time_complexity_centralized_TL}, we see that under suitable conditions the TL-based approach performs better than the source domain sample-based agnostic approach. The following inferences are drawn:
\begin{enumerate}[(1)]
\item The minimum user density to achieve a finite delay is reduced by a positive offset of $2 \epsilon_{pq}^2 m$. In fact, for $m > \frac{\log{\left(\frac{2N}{\delta}\right)}}{2(\bar \epsilon - \norm{\mathcal{P}-\mathcal{Q}}_\mathcal{H})}$, a finite delay can be achieved for all user density which provides significant advantage.

\item The finite delay achieved is smaller compared to source domain sample-based agnostic approach for large enough source samples, and the distributions are ``close.'' This is made precise in the following proposition.
\end{enumerate}

\begin{prop}
For any $\epsilon>0$ and $\delta \in [0,1]$, the TL-based approach performs better than the source sample-based agnostic approach provided the number of source samples satisfies
\beq
m \geq \frac{1}{2 \epsilon_{pq}^2} \left[\log\left(\frac{2N}{\delta}\right) - F\right]^+,
\eeq
and the distributions satisfy the following condition
\beq
\norm{\mathcal{P}-\mathcal{Q}}_{\mathcal{H}} < \frac{\epsilon R_0}{2 B \lambda_u \pi \gamma^2  N},
\eeq
where
\beq
F \triangleq \lambda_u \pi R^2 \left(1 - \exp\left\{\frac{1 - e^{-2 \bar{\epsilon}^2}}{1 - e^{-2 {\epsilon}_{pq}^2}}\right\} \left(1 - \mathcal{L}\right) \right)
\eeq
and $\mathcal{L} \triangleq \frac{1}{\lambda_u \pi R^2} \log\left(\frac{2N}{\delta}\right)$.
\label{prop:TL_cache}
\end{prop}

\section{Concluding Remarks}\label{sec:conclude}
This paper considered the problem of caching in a distributed heterogenous cellular network, when the popularity profiles of cached data files were unknown. A heterogenous network was considered where BS, SBSs and users were distributed according to independent PPPs. SBSs stored high data-rate content, which could be downloaded directly by one of the users from a SBS in the vicinity. A metric was proposed that captured the offloading-loss, which was used to optimize a random caching strategy. This metric was shown to be a function of the popularity profile (assumed unknown).

The popularity profile was estimated at BS using the available instantaneous demands from users in a time interval $[0,\tau]$. We showed that a waiting time $\tau$ to achieve an $\epsilon>0$ difference between the achieved cost and the optimal cost was finite, provided the user density was greater than a threshold. In this case, $\tau$ was shown to scale as $N^2$, where $N$ was the support of the popularity profile. A TL-based approach was proposed to estimate the popularity profile, which was then used to compute the empirical cost. A condition was derived under which the TL-based approach performed better than the random caching strategy.

Although TL-based approach performs better, the improvement is not without a price. The error that is achieved in \eqref{eq:epsilon_error_bound} depends on $\norm{\mathcal{P}-\mathcal{Q}}_\mathcal{H}$, suggesting that lower the distance between the two distributions better the TL scheme performs. If $\norm{\mathcal{P}-\mathcal{Q}}_\mathcal{H} = 0$, the TL-based approach performs significantly better than the other proposed scheme. However, from Proposition \ref{prop:TL_cache}, the benefits of using target domain samples can only be realized with the knowledge of the distance $\norm{\mathcal{P}-\mathcal{Q}}_\mathcal{H}$. In practice, an accurate estimate of this distance is difficult to obtain. However, there could be hope of finding an \emph{efficient estimate} of $\norm{\mathcal{P}-\mathcal{Q}}_\mathcal{H}$, and this is relegated to future work.

Other potential avenues for future research for the problem setup considered in this paper include:
\begin{enumerate}[(1)]
\item The popularity profile was assumed to be constant across users and stationary in time, which, in practice, may not be justifiable. Analysis of the performance of caching strategies by relaxing these assumptions merits investigation.
\item Characterizing the performance of TL-based caching when the popularity profile is estimated using linear combinations of source domain and target domain samples is another interesting topic for research.
\end{enumerate}

\bibliographystyle{IEEEtran}
\bibliography{IEEEabrv,mybibdata1}

\begin{thebibliography}{10}
\providecommand{\url}[1]{#1}
\csname url@samestyle\endcsname
\providecommand{\newblock}{\relax}
\providecommand{\bibinfo}[2]{#2}
\providecommand{\BIBentrySTDinterwordspacing}{\spaceskip=0pt\relax}
\providecommand{\BIBentryALTinterwordstretchfactor}{4}
\providecommand{\BIBentryALTinterwordspacing}{\spaceskip=\fontdimen2\font plus
\BIBentryALTinterwordstretchfactor\fontdimen3\font minus
  \fontdimen4\font\relax}
\providecommand{\BIBforeignlanguage}[2]{{%
\expandafter\ifx\csname l@#1\endcsname\relax
\typeout{** WARNING: IEEEtran.bst: No hyphenation pattern has been}%
\typeout{** loaded for the language `#1'. Using the pattern for}%
\typeout{** the default language instead.}%
\else
\language=\csname l@#1\endcsname
\fi
#2}}
\providecommand{\BIBdecl}{\relax}
\BIBdecl

\bibitem{Akyildiz2004}
I.~F. Akyildiz, J.~Xie, and S.~Mohanty, ``A survey of mobility management in
  next-generation \mbox{all-IP-based} wireless systems,'' \emph{IEEE Trans.
  Wireless Commun.}, vol.~11, no.~4, pp. 16--28, Aug. 2004.

\bibitem{Chou2014}
S.-F. Chou, T.-C. Chiu, Y.-J. Yu, and A.-C. Pang, ``Mobile small cell
  deployment for next generation cellular networks,'' in \emph{Proc. IEEE
  Global Commun. Conf.}, Dec. 2014, pp. 4852--4857.

\bibitem{Bennis2013}
M.~Bennis, M.~Simsek, A.~Czylwik, W.~Saad, S.~Valentin, and M.~Debbah, ``When
  cellular meets \mbox{WiFi} in wireless small cell networks,'' \emph{IEEE
  Commun. Magazine}, vol.~51, no.~6, pp. 44--50, Jun. 2013.

\bibitem{Kim2014}
J.~Kim, C.~Jeong, H.~Yu, and J.~Park, ``Areal capacity limit on the growth of
  small cell density in heterogeneous networks,'' in \emph{Proc. IEEE Global
  Commun. Conf.}, Dec. 2014, pp. 4263--4268.

\bibitem{Lin2003}
Y.-B. Lin, W.-R. Lai, and J.-J. Chen, ``Effects of cache mechanism on wireless
  data access,'' \emph{IEEE Trans. Wireless Commun.}, vol.~2, no.~6, pp.
  1247--1258, Nov. 2003.

\bibitem{Bastug2014}
E.~Bastug, M.~Bennis, and M.~Debbah, ``Living on the edge: The role of
  proactive caching in \mbox{5G} wireless networks,'' \emph{IEEE Commun.
  Magazine}, vol.~52, no.~8, pp. 82--89, Aug. 2014.

\bibitem{Niesen2012}
U.~Niesen, D.~Shah, and G.~W. Wornell, ``Caching in wireless networks,''
  \emph{IEEE Trans. Inf. Theory}, vol.~58, no.~10, pp. 6524--6540, Oct. 2012.

\bibitem{Sadjadpour2014}
H.~R. Sadjadpour, ``A new design for \mbox{Information Centric Networks},'' in
  \emph{Proc. Conf. Inf. Sciences and Syst.}, Mar. 2014, pp. 1--6.

\bibitem{Ali_Niessen_ISIT_may2014}
M.~A. Maddah-Ali and U.~Niesen, ``Fundamental limits of caching,'' \emph{IEEE
  Trans. Inf. Theory}, vol.~60, no.~5, pp. 2856--2867, May 2014.

\bibitem{Maddah-Ali2014a}
------, ``Decentralized coded caching attains order-optimal memory-rate
  tradeoff,'' in \emph{Proc. Allerton Conf. Commun., Control, Comp.}, Oct.
  2013, pp. 421--427, also to appear in IEEE/ACM Trans. Networking.

\bibitem{Niessen_Ali_INFOCOM_April2014}
U.~Niesen, C.~Beilken, and M.~A. Maddah-Ali, ``Coded caching with nonuniform
  demands,'' in \emph{Proc. IEEE Conf. Computer Commun. Workshop}, Apr. 2014,
  pp. 221--226.

\bibitem{Niessen_Ali_arxiv_2014}
\BIBentryALTinterwordspacing
U.~Niesen and M.~A.~M. Ali, ``Coded caching for delay-sensitive content,''
  2014, tech. report. [Online]. Available: \url{http://arxiv.org/abs/1407.4489}
\BIBentrySTDinterwordspacing

\bibitem{Pedarsani_Ali_INFOCOM_ICC_June2014}
R.~Pedarsani, M.~A. Maddah-Ali, and U.~Niesen, ``Online coded caching,'' in
  \emph{Proc. IEEE Int. Conf. Commun.}, Jun. 2014, pp. 1878--1883.

\bibitem{Karamchandani_Niesen_Ali_Diggavi_arxiv_2014}
\BIBentryALTinterwordspacing
N.~Karamchandani, U.~Niesen, M.~A. Maddah{-}Ali, and S.~N. Diggavi,
  ``Hierarchical coded caching,'' 2014, submitted to IEEE Trans. on Inf.
  Theory. [Online]. Available: \url{http://arxiv.org/abs/1403.7007}
\BIBentrySTDinterwordspacing

\bibitem{Guan2014}
Y.~Guan, Y.~Xiao, H.~Feng, C.-C. Shen, and L.~J. Cimini, ``\mbox{MobiCacher}:
  Mobility-aware content caching in small-cell networks,'' in \emph{Proc. IEEE
  Global Commun. Conf.}, Dec. 2014, pp. 4537--4542.

\bibitem{Ji_Caire_Molisch_ISIT_July2013}
M.~Ji, G.~Caire, and A.~F. Molisch, ``Optimal throughput-outage trade-off in
  wireless one-hop caching networks,'' in \emph{Proc. IEEE Int. Symp. Inf.
  Theory}, Jul. 2013, pp. 1461--1465.

\bibitem{Golrezaei_Dimakis_Molisch_ISIT_July2013}
N.~Golrezaei, A.~G. Dimakis, and A.~F. Molisch, ``Wireless device-to-device
  communication with distributed caching,'' in \emph{Proc. IEEE Int. Symp. Inf.
  Theory}, Jul. 2012, pp. 2781--2785.

\bibitem{Ji_Caire_Molisch_ITW_Sep2013}
M.~Ji, G.~Caire, and A.~F. Molisch, ``Fundamental limits of distributed caching
  in \mbox{D2D} wireless networks,'' in \emph{Proc. IEEE Inf. Theory Workshop},
  Sep. 2013, pp. 1--5.

\bibitem{Golrezaei_Shanmugam_Dimakis_Molisch_TIT_Dec2013}
N.~Golrezaei, K.~Shanmugam, A.~Dimakis, A.~Molisch, and G.~Caire, ``Femto
  caching: Wireless video content delivery through distributed caching
  helpers,'' \emph{IEEE Trans. Inf. Theory}, vol.~59, no.~12, pp. 8402--8413,
  Dec. 2013.

\bibitem{MingyueJi_Caire_Molisch_arxiv_2014}
\BIBentryALTinterwordspacing
M.~Ji, G.~Caire, and A.~F. Molisch, ``Fundamental limits of caching in wireless
  {D2D} networks,'' 2014, submitted to IEEE Trans. on Inf. Theory. [Online].
  Available: \url{http://arxiv.org/abs/1405.5336}
\BIBentrySTDinterwordspacing

\bibitem{Blasco_Gunduz_ICC_June2014}
P.~Blasco and D.~Gunduz, ``Learning-based optimization of cache content in a
  small cell base station,'' in \emph{Proc. IEEE Int. Conf. Commun.}, Jun.
  2014, pp. 1897--1903.

\bibitem{Blasco_Gunduz_ISIT_June2014}
------, ``Multi-armed bandit optimization of cache content in wireless
  infostation networks,'' in \emph{Proc. IEEE Int. Symp. Inf. Theory}, Jun.
  2014, pp. 51--55.

\bibitem{Sengupta_Amura_Tandom_Buehrer_Clancy_ISWCS_2014}
A.~Sengupta, S.~Amuru, R.~Tandon, R.~M. Buehrer, and T.~C. Clancy, ``Learning
  distributed caching strategies in small cell networks,'' in \emph{Proc. IEEE
  Int. Symp. Wireless Commun. Syst.}, 2014.

\bibitem{FBaccelli_SpringerJS_1997}
F.~Baccelli, M.~Klein, M.~Lebourges, and S.~Zuyev, ``Stochastic geometry and
  architecture of communication networks,'' \emph{J. Telecom. Syst.}, vol.~7,
  no.~1, pp. 209--227, 1997.

\bibitem{Bastug_Bennis_Debbah_KuVS_2014}
E.~Bastug, M.~Bennis, and M.~Debbah, ``Anticipatory caching in small cell
  networks: A transfer learning approach,'' in \emph{Workshop on Anticipatory
  Networks}, Germany, Sep. 2014.

\bibitem{Prob_theory_PR_Devroye_Gyorfi_Lugosi}
L.~Devroye, L.~Gyorfi, and G.~Lugosi, \emph{A Probability Theory of Pattern
  Recognition}.\hskip 1em plus 0.5em minus 0.4em\relax Springer, 2014.

\end{thebibliography}

%\section{Appendix}
%\subsection{Useful Lemma (see \cite{Prob_theory_PR_Devroye_Gyorfi_Lugosi})} \label{app:hoeffdings}
%\begin{lem} \label{lemm:hoeffdings}
%(Hoeffdings) Let $X_1,X_2,\ldots,X_n$ be independent random variables. Further, $X_i \in [a_i,b_i]$, $i=1,\ldots,n$. Then, for any $\epsilon >0$
%\beq
%\Pr\left\{\abs{S_n - \mathbb{E} S_n} > \epsilon\right\} \leq  2 \exp\left\{-\frac{2 n^2 \epsilon^2}{\sum_{i=1}^n (a_i - b_i)^2}\right\},
%\eeq
%where $S_n := \sum_{i=1}^n X_i$.
%\end{lem}

\appendices

\section{Proof of Theorem \ref{thm_mean_throughput}} \label{app:througput_derivation}
Consider the first term in \eqref{eq:metric}
\beqa \label{eq:mean_througput_expression_1term}
\mathbb{E} \mathbf{1} \{f_i \notin \mathcal{N}_u\} \hspace{-2mm}&=&\hspace{-2mm} \mathbb{E}_{n_s} \Pr\{f_i \notin \mathcal{N}_u \vert \abs{\mathcal{N}_u} = n_s\} \nonumber\\
\hspace{-2mm}&\stackrel{(a)}{=}&\hspace{-2mm} \mathbb{E}_{n_s}\left[\Pr\{f_i \notin a,\text{ for some } a \in \mathcal{N}_u\}\right]^{n_s} \nonumber\\
\hspace{-2mm}&\stackrel{(b)}{=}&\hspace{-2mm} \mathbb{E}_{n_s}(1-\pi_i)^{n_s M} \nonumber\\
\hspace{-2mm}&\stackrel{(c)}{=}&\hspace{-2mm} \mathbb{E} \sum_{j=0}^\infty (1-\pi_i)^{j M} e^{\{-\lambda_s \pi \gamma^2\}} \frac{(\lambda_s \pi \gamma^2)^j}{{j !}} \nonumber \\
%\hspace{-2mm}&=&\hspace{-2mm} \mathbb{E} \exp\left\{-\lambda_s \pi \gamma^2 \left[1-(1-\pi_i)^M\right]\right\} \nonumber \\
\hspace{-2mm}&=&\hspace{-2mm} \sum_{i=1}^N  \exp\left\{-\mathcal{U}\right\}  p_i, \label{eq:mean_througput_expression_1term}
\eeqa
where $\mathcal{U}\triangleq\lambda_s \pi \gamma^2 \left[1-(1-\pi_i)^M\right]$. In the above exposition, $(a)$ follows from the fact that the random caching proposed in this paper is independent across users, $(b)$ follows since the cache size is fixed to be $M$, and $(c)$ follows due to the fact that $n_s$ is a  PPP with mean $\lambda_s \pi \gamma^2$ since $n_s$ is the number of SBSs in a circular area of radius $\gamma$. $\blacksquare$
%Now, consider a part of the second term, i.e., $\mathcal{J}:= \mathbb{E} \mathbf{1} \{u \text{ not } T \text{-feasible}\} $ in  \eqref{eq:mean_througput_expression}:
%\beqa
%\mathcal{J} &=& \mathbb{E}_{\Phi_u} \Pr \{u \text{ not } T \text{-feasible} | k \text{ users in } \mathbb{B}(0,\gamma)\} \nonumber\\
%&=& \mathbb{E}_{\Phi_u} \Pr \{q^{kT} | k \text{ users in } \mathbb{B}(0,\gamma)\} \nonumber\\
%&=& \sum_{k=0}^\infty q^{kT} \exp\{-\lambda_u \pi \gamma^2 \} \left[\frac{(\lambda_u \pi \gamma^2)^k}{k!}\right] \nonumber\\
%&=& \exp\{-\lambda_u \pi \gamma^2(1-q^T)\}. \label{eq:throughput_second_term}
%\eeqa
%Using \eqref{eq:mean_througput_expression_1term} and \eqref{eq:throughput_second_term} in \eqref{eq:mean_througput_expression}, the desired expression of the theorem is obtained.
%\newpage
\section{Proof of Theorem \ref{thm:time_complexity_centralized}} \label{app:time_complexity_centralized}
First, for any $\epsilon>0$, the following inequality is proved \cite{Prob_theory_PR_Devroye_Gyorfi_Lugosi}:
\beq \label{eq:prob_risk_bound_1}
\Pr\{\hat{\mathcal{T}}^* \geq \mathcal{T}^* + \epsilon\} \leq \Pr\left\{2 \sup_{\mathbf{1} \succeq \Pi \succeq \mathbf{0}:\mathbf{1}^T \Pi = 1} \abs{\Delta \mathcal{T}} > \epsilon\}\right\},
\eeq
where $\Delta \mathcal{T}\triangleq \mathcal{T}(\Pi,\hat{\mathcal{P}}) - \mathcal{T}(\Pi,\mathcal{P})$. We have,%\footnote{The description of the constraint in the following set of inequalities is omitted for brevity.}
\beqa
\hat{\mathcal{T}}^*- \mathcal{T}^*   &=& \hat{\mathcal{T}}^* - \inf_{\Pi} \mathcal{T}(\Pi,{\mathcal{P}}) \nonumber\\
%&=&   \hat{\mathcal{T}^*} -  \hat{\mathcal{T}}  +  \hat{\mathcal{T}}  -    \inf_{\Pi} \mathcal{T}(\Pi,{\mathcal{P}}) \nonumber \\
&\leq&   \hat{\mathcal{T}^*} -  \hat{\mathcal{T}}  + \sup_{\Pi} \abs{\mathcal{T}(\Pi,\hat{\mathcal{P}}) - \mathcal{T}(\Pi,{\mathcal{P}})} \nonumber\\
&\leq& 2\sup_{\Pi} \abs{\mathcal{T}(\Pi,\hat{\mathcal{P}}) - \mathcal{T}(\Pi,{\mathcal{P}})},
\eeqa
where $\hat{\mathcal{T}}\triangleq\mathcal{T}(\Pi,\hat{\mathcal{P}})$, thus proving \eqref{eq:prob_risk_bound_1}. Now, consider the right hand side of \eqref{eq:prob_risk_bound_1} after substituting for $\mathcal{T}(\Pi,\hat{\mathcal{P}})$ and $\mathcal{T}(\Pi,{\mathcal{P}})$ from \eqref{eq:mean_througput_expression} to get $\Pr \left\{ \sup_\Pi  \abs{\sum_{i=1}^N g(\pi_i) (\hat{p}_i - p_i)} > \tilde{\epsilon} \right\} $, which can be upper bounded as follows:
\beqa
\Pr \left\{ \sup_\Pi {\sum_{i=1}^N g(\pi_i) \hat{\delta}_p} > \tilde{\epsilon} \right\}
\hspace{-2mm} &\leq& \hspace{-2mm} \Pr \left\{{ \max_{i=1,2,\ldots,N} \hat{\delta}_p} > \bar{\epsilon} \right\} \nonumber \\
\hspace{-2mm}&\leq& \hspace{-2mm}\sum_{i=1}^N \Pr \left\{{ \hat{\delta}_p} > \bar{\epsilon} \right\} \nonumber \\
\hspace{-2mm}&\leq&\hspace{-2mm} 2N \mathbb{E}\left[\exp\left\{-  {2 {\bar{\epsilon}}^2 n_p} \right\}\right], \label{eq:hoeffdings_baseline}
\eeqa
where $\hat{\delta}_p \triangleq \abs{\hat{p}_i - p_i}$, $\tilde \epsilon \triangleq  \frac{R_0 \epsilon}{2 B}$, $\bar \epsilon \triangleq \frac{\tilde{\epsilon}}{\sup_\Pi \left\{\sum_{i=1}^N g(\pi_i)\right\} }$, and $g(\pi_i)\triangleq \exp\{-\lambda_u \pi \gamma^2(1-\pi_i)^M\}$. The last inequality above follows by applying Hoeffdings inequality (see \cite{Prob_theory_PR_Devroye_Gyorfi_Lugosi}), since the estimator $\hat{\mathcal{P}}$ is unbiased and $\pi_i$, $\pi \in [0,1]$. The expectation in \eqref{eq:hoeffdings_baseline} is with respect to $n_p$. Conditioned on the number of users $n_R$ in the coverage area of BS, $n_p$ is a Poisson distributed random variable with density $n_R \lambda_r \tau$. We let $\bar g \triangleq \left({2 \bar{\epsilon}^2 k } + \lambda_r n_R \tau\right)$, and $g^* \triangleq \left( 1 - \exp\left\{-{2 \bar{\epsilon}^2 }\right\}\right)$. Using this, we get
\beq
2 N \mathbb{E} \sum_{n=0}^\infty \exp\left\{-\bar g\right\} \frac{(\lambda_r n_R \tau)^n}{n!} = 2 N \mathbb{E}_{n_R} \exp\{-\lambda_r n_R \tau g^*\}.
\eeq
This can be further simplified as
\beqa \label{eq:error_bound}
&2 N \sum_{k=0}^\infty  \exp\{-\lambda_r k \tau  g^*\} \exp\{-\lambda_u \pi R^2\} \frac{(\lambda_u \pi R^2)^k}{k!} = \nonumber \\ &2 N\exp\{-\lambda_u \pi R^2 \left(1- \exp\left\{-\lambda_r \tau g^*\right\} \right).
\eeqa
Now, it is easy to see that $\Pr\left\{ \sup_{\Pi} \abs{\Delta \mathcal{T}} > \frac{\epsilon}{2}\right\} \leq \delta$ if the expression in \eqref{eq:error_bound} is upper bounded by $\delta$, which implies that
\beq
\tau \geq \frac{1}{\lambda_r g^*} \log \left(\frac{1}{1- \frac{1}{\lambda_u \pi R^2} \log \frac{2 N}{\delta}}\right)
\eeq
provided $\lambda_u > \frac{1}{\pi R^2} \log \frac{2 N}{\delta}$,  otherwise $\tau = \infty$. This completes the proof of the theorem. $\blacksquare$

\section{Proof of Theorem \ref{thm:time_complexity_centralized_TL}} \label{app:TL_time_complexity}
As in the proof of Theorem \ref{thm:time_complexity_centralized}, it is easy to see that
\beq
\Pr\{\hat{\mathcal{T}}^* \geq \mathcal{T}^* + \epsilon\} \leq \Pr \left\{ \sup_{1\leq i \leq N} \abs{\hat{p}_i^{(tl)} - p_i} > \bar{\epsilon} \right\},
\eeq
where $\bar \epsilon \triangleq \frac{R_0 {\epsilon}}{2 B \sup_\Pi \left\{\sum_{i=1}^N g(\pi_i)\right\} }$, and $g(\pi_i)\triangleq\exp\{-\lambda_u \pi \gamma^2(1-\pi_i)^M\}$. Denote by $n_{p}$ the total number of requests in the coverage area of BS. As mentioned previously, conditioned on the number of users $n_R$ in the coverage area of BS, $n_{p}$ is a Poisson distributed random variable with density $n_R \lambda_r$. Further, $\mathbb{E}\left\{ \hat{p}_i^{(tl)} | n_p\right\} = \frac{n_{p}}{n_{p} + m}p_i + \frac{m}{n_{p} + m}q_i$. Using this, we can write
\beqa
&&\Pr \left\{ \sup_{1\leq i \leq N} \abs{\hat{p}_i^{(tl)}  - \mathbb{E} \hat{p}_i^{(tl)} + \mathbb{E} \hat{p}_i^{(tl)} - p_i} > \bar{\epsilon} \right\} \nonumber\\
&\leq& \Pr \left\{ \sup_{1\leq i \leq N} \abs{\hat{p}_i^{(tl)}  - \mathbb{E} \hat{p}_i^{(tl)}} + \abs{\mathbb{E} \hat{p}_i^{(tl)} - p_i} > \bar{\epsilon} \right\}\\
&\leq& \Pr \left\{ \sup_{1\leq i \leq N} \abs{\hat{p}_i^{(tl)}  - \mathbb{E} \hat{p}_i^{(tl)}}  > \bar{\epsilon} -  \sup_{i\in[1,N]} \abs{\mathbb{E} \hat{p}_i^{(tl)} - p_i}\right\} \nonumber\\
&\leq& \Pr \left\{ \sup_{1\leq i \leq N} \abs{\hat{p}_i^{(tl)}  - \mathbb{E} \hat{p}_i^{(tl)}}  > \bar{\epsilon} -  \frac{m}{n_{p} + m} \norm{\mathcal{P} - \mathcal{Q}}_{\mathcal{H}} \right\} \nonumber\\
&\leq& \mathbb{E}_{n_{p}}\Pr \left\{ \sup_{1\leq i \leq N} \abs{\hat{p}_i^{(tl)}  - \mathbb{E} \hat{p}_i^{(tl)}}  > \bar{\epsilon} -   \norm{\mathcal{P} - \mathcal{Q}}_{\mathcal{H}} \left | \right. n_{p} \right\}\nonumber \\
&\leq& N \mathbb{E}_{n_{p}}\Pr \left\{\abs{\hat{p}_i^{(tl)}  - \mathbb{E} \hat{p}_i^{(tl)}}  > \bar{\epsilon} -   \norm{\mathcal{P} - \mathcal{Q}}_{\mathcal{H}} \left | \right. n_{p} \right\},\nonumber
\eeqa
provided $\bar \epsilon >  \norm{\mathcal{P} - \mathcal{Q}}_{\mathcal{H}}$. Now, employing Hoeffding's inequality, we can write
\beqa
 && 2 N \mathbb{E}_{n_{p}} \exp\left\{-2 \epsilon_{pq}^2 (n_{p} + m) \right\}  \nonumber \\ &=& 2N \mathbb{E}_{n_R} \exp\left\{-(2 \epsilon_{pq}^2 m +\lambda_r n_R \tau)\right\} \sum_{k=0}^\infty \frac{a^k}{k!},\nonumber \\
 &=& 2 N \exp\left\{-2 \epsilon_{pq}^2 m\right\} \mathbb{E}_{n_R} \exp\{-\bar{g}_{pq}\} \nonumber \\
 &=& 2 N \exp\left\{-2 \epsilon_{pq}^2 m\right\} \exp\{-\lambda_u \pi R^2\} \times \nonumber\\
 && \sum_{l=0}^\infty \exp\{- \lambda_r l \tau(1-\exp\{-2 \epsilon_{pq}^2\})\} \frac{(\lambda_u \pi R^2)^l}{l!}, \nonumber
 \eeqa
where $a \triangleq \lambda_r n_R \tau \exp\left\{-2 \epsilon_{pq}^2 \right\} $, $\epsilon_{pq} \triangleq \bar \epsilon - \norm{\mathcal{P} - \mathcal{Q}}_{\mathcal{H}}$ and $\bar{g}_{pq}\triangleq \lambda_r n_R \tau(1-\exp\{-2\epsilon_{pq}^2\}) $. Therefore, we have
\begin{eqnarray}
\nonumber && 2 N \mathbb{E}_{n_{p}} \exp\left\{-2 \epsilon_{pq}^2 (n_{p} + m) \right\}\\ &=& 2 \exp\left\{-2 \epsilon_{pq}^2 m\right\} \exp\{-\lambda_u \pi R^2 t\},
\end{eqnarray}
where $t \triangleq \left(1 - \exp\{-\lambda_r  \tau\left(1 - \exp\{-2 \epsilon_{pq}^2\} \right)\}\right)$. The above is at most $\delta >0$ if
\beq
\tau \geq \frac{1}{\lambda_r (1-e^{-2 \epsilon_{pq}^2})} \log \left(\frac{1}{1 - \frac{1}{\lambda_u \pi R^2} \left(\log \frac{2N}{\delta} - 2 \epsilon_{pq}^2 m\right)} \right) \nonumber
\eeq
provided
\beq
\lambda_u >  \frac{1}{\pi R^2} \left(\log \frac{2N}{\delta} - 2 \epsilon_{pq}^2 m\right),
\eeq
otherwise $\tau = \infty$. This completes the proof of Theorem \ref{thm:time_complexity_centralized_TL}. $\blacksquare$
%\mathbf{1} \succeq \Pi \succeq \mathbf{0}:\mathbf{1}^T \Pi = 1
%\begin{thebibliography}
%\bibliographystyle{IEEEtran.bst}
%\end{thebibliography}
%\bibliography{mybib}
%##################################################################################################################################

\end{document}